\documentclass[conference]{IEEEtran}
\IEEEoverridecommandlockouts

\usepackage{subfigure}

\usepackage{graphicx}

\ifCLASSINFOpdf
   \usepackage[pdftex]{graphicx}
\else
\fi
\usepackage[cmex10]{amsmath}
\usepackage{amssymb}  % assumes amsmath package installed
\usepackage{amsfonts}

\usepackage{array}
\begin{document}
\title{Quantum filter for a class of non-Markovian quantum systems}

%\author{%\IEEEauthorblockN{Shi-Bei~Xue}
%%\IEEEauthorblockA{School of Electrical and\\Computer Engineering\\
%%Georgia Institute of Technology\\
%%Atlanta, Georgia 30332--0250\\
%%Email: http://www.michaelshell.org/contact.html}
%\and
%\IEEEauthorblockN{Matthew~R.~James}
%%\IEEEauthorblockA{Twentieth Century Fox\\
%%Springfield, USA\\
%%Email: homer@thesimpsons.com}
%\and
%
%%\IEEEauthorblockA{Starfleet Academy\\
%%San Francisco, California 96678-2391\\
%%Telephone: (800) 555--1212\\
%%Fax: (888) 555--1212}
%\IEEEauthorblockN{Alireza Shibani}
%\IEEEauthorblockN{Valery Ugrinovskii\\ and Ian~R.~Petersen}
%}

\author{Shibei~Xue, Matthew~R.~James, Alireza Shabani, Valery Ugrinovskii, and~Ian~R.~Petersen
\thanks{
This research was supported under Australian Research Council¡¯s Discovery
Projects and Laureate Fellowships funding schemes (Projects DP140101779 and FL110100020).}
\thanks{S. Xue, V. Ugrinovskii and I. R. Petersen are with the School of Information Technology and Electrical Engineering, University of New South Wales Canberra at the Australian Defence Force Academy, Canberra, ACT 2600, Australia (e-mail: xueshibei@gmail.com; v.ugrinovskii@gmail.com; i.r.petersen@gmail.com).}% <-this % stops a space
\thanks{M. R. James is with the ARC Centre for Quantum Computation and
Communication Technology, Research School of Engineering, Australian
National University, Canberra, ACT 0200, Australia (e-mail: Matthew.James@anu.edu.au).}
\thanks{A. Shabani is a research scientist at Google Quantum Artificial Intelligence Lab, Google, 340 Main St. Venice, CA 90291, U.S.A. (e-mail: shabani@google.com).}% <-this % stops a space
}

%Telephone: (800) 555--1212, Fax: (888) 555--1212}


\maketitle

\begin{abstract}
In this paper we present a Markovian representation approach to constructing  quantum filters for a class of non-Markovian quantum systems disturbed by Lorentzian noise. An ancillary system is introduced to convert white noise into Lorentzian noise which is injected into a principal system via a direct interaction. The resulting dynamics of the principal system are non-Markovian, which are driven by the Lorentzian noise. By probing the principal system,  a quantum filter for the augmented system can be derived from standard theory, where the conditional  state of the principal system can be obtained by tracing out the ancillary system. An example is provided  to illustrate  the non-Markovian dynamics of the principal system.
\end{abstract}


% \fi
\IEEEpeerreviewmaketitle

\section{Introduction}
The rapid growth of quantum information technology greatly boosts the development of quantum control theory~\cite{bouten} which in turn makes such technology more reliable. For example, quantum filtering~\cite{bouten} and quantum feedback control~\cite{hinfinity,Wiseman1994,HarocheNAT2011} have been applied in stabilizing quantum information carriers~\cite{Mirrahimi}, in systematically designing photonic computation circuits~\cite{Qm}, and in enhancing the performance of quantum metrology~\cite{Altafini2012}.

The quantum plants that most existing works are concerned with are Markovian, and are disturbed by white noise only~\cite{Breuer,Gardiner}. However, many problems of interest involve complicated environmental influences, e.g., colored noise~\cite{Xue2011,XuePRA2012}, which if not taken into account may lead to degraded performance of estimation and control schemes.
Similar modelling issues arise for classical systems~\cite{Hanggi}, and a common approach in control engineering to account for the non-Markovian effects of colored noise is to augment the system with a whitening filter~\cite{KS72}.

In this paper we consider non-Markovian quantum systems by representing them in a larger  Markovian system framework.
If we assume that the principal system of interest is defined on a Hilbert space $\mathfrak{h}$, and the noise on a Fock space $\mathfrak{F}$,  we introduce an
ancillary system defined on the Hilbert space $\mathfrak{h}_0$ converting white noise into colored noise to  model the internal modes of the non-Markovian environment, whose structure determines the spectrum of the colored noise. Then the augmented system has a Markovian evolution on the augmented Hilbert space $\mathfrak{h}\otimes\mathfrak{h}_0\otimes\mathfrak{F}$. Such an approach was proposed in~\cite{PhysRevA.50.3650} to model a non-Markovian system and was named as a pseudo-mode method later~\cite{PhysRevA.80.012104}. This approach is also applied to model energy transfer process in photosynthetic complexes~\cite{JCP}.
Similarly, the augmented system model can be realized by a quantum collision model, where the spectrum of the noise is implicitly determined by the ancillary system~\cite{PhysRevA.88.032115}. Also, a hierarchy equation approach has been adopted to describe the dynamics of non-Markovian quantum systems~\cite{PhysRevA.85.062323}, where parts of the equations describe the pseudo-mode dynamics. This has been applied to the indirect continuous measurement of a non-Markovian quantum system~\cite{shabani2014}. However, this pseudo-mode approach has not been systematically described so as to be compatible with quantum control theory, e.g., quantum filtering theory.

%The previous work~\cite{xue2013} has shown a Green's function of the non-Markovian system with Lorentzian noise is in a second order. Hence, it is possible to represent the non-Markovian system by two coupled first-order Markovian quantum systems.

For concreteness,
in this paper, we assume that the noise has a Lorentzian spectrum  and employ a quantum stochastic differential equation (QSDE) approach to represent the colored noise via an internal mode of the environment.
Due to the prevalence of Lorentzian noise in some solid-state systems~\cite{WeiMinPRL2012,Tu2008}, it is useful to model Lorentzian noise in quantum control applications. This colored noise is injected into the principal system of interest via a direct interaction such that the dynamics of the principal system can be described by a quantum stochastic integral differential equation (QSIDE), i.e., a non-Markovian Langevin equation. In addition, the augmented model of the non-Markovian quantum system can be conveniently described by an $(S,L,H)$ description in an extended Hilbert space which is compatible with quantum filtering theory.  The quantum filter for the non-Markovian quantum system can be constructed by injecting a probing field into the principal system. Due to the output field  satisfying a non-demolition condition, the augmented system state can be estimated by the filter, with which the non-Markovian dynamics of the principal system can be obtained by tracing out the ancillary system.  We note however that more general noise spectra can be considered in an analogous manner (more details will be presented in a future paper).

The paper is organized as follows. In Section~\ref{sec2}, based on a Markovian quantum system model, we introduce a colored noise model to model Lorentzian noise. In Section~\ref{sec3}, we show that the principal system disturbed by the noise from the noise model satisfies a QSIDE. In Section~\ref{sec4}, to observe such a system, a probing field is applied to the principal system, whose output is observed, and then we can construct a quantum filter. A possible experimental realization is discussed in Section~\ref{sec5}. Conclusions are drawn in Section~\ref{sec6}.
\section{Dynamics of the ancillary system driven by white noise}\label{sec2}
\subsection{Review of Markovian quantum systems}
%\subsubsection{Definition of Markovian quantum systems}
%\cite{HP84}
\subsubsection{Hamiltonian}
Consider a quantum system defined on a Hilbert space $\mathfrak{h}$ interacting with an electromagnetic field defined on the Boson Fock space $\mathfrak{F}$ over $L^2(\mathbb{R}_+)$. For example, this system may be represented by an optical mode in a cavity interacting with a probing field. %as shown in Fig.~\ref{f1}.
The unitary dynamics of the total system on the space $\mathfrak{h}\otimes\mathfrak{F}$ is defined by the Hamiltonian
\begin{equation}\label{1}
  H=H_S+H_I+H_F
\end{equation}
with the total system evolution operator $ \Lambda_t={\rm exp}(-{\rm i}Ht)$,
where $H_S$ is the system Hamiltonian describing its free evolution on the space $\mathfrak{h}$ and hereafter we set Planck's constant $\hbar=1$. The free evolution of the field on the space $\mathfrak{F}$ is described by a unitary operator $ \Theta_t={\rm exp}(-{\rm i}H_F t)$,
which is determined by a field Hamiltonian $H_F=\int_{-\infty}^{+\infty}\omega b^\dagger(\omega)b(\omega){\rm d}\omega$
with boson annihilation (creation) operator $b(\omega)$ ($b^\dagger(\omega)$) defined on the space $\mathfrak{F}$ satisfying a delta commutation relation $[b(\omega),b^\dagger(\omega')]=\delta(\omega-\omega')$.
The interaction Hamiltonian $H_I$ can be expressed as $H_I={\rm i}\int_{-\infty}^{+\infty}(b^\dagger(\omega)L-L^\dagger b(\omega)){\rm d}\omega$,
where the coupling operator $L$ acting on the space $\mathfrak{h}$ only is expressed as a product between a decoherence rate $\sqrt{\gamma}\in\mathbb{R}_+$ and a system operator.
\subsubsection{Dynamical Equation}
In the interaction picture, the effective Hamiltonian can be obtained as
\begin{equation}\label{7}
  H_{\rm eff}(t)=\Theta^\dagger_t H \Theta_t=H_S+{\rm i}(b^\dagger(t)L-L^\dagger b(t)),
\end{equation}
where the field is defined as
\begin{equation}\label{5}
  b(t)=\frac{1}{\sqrt{2\pi}}\int_{-\infty}^{+\infty}b(\omega)e^{-{\rm i}\omega t}{\rm d}\omega
\end{equation}
satisfying the delta commutation relations
\begin{equation}\label{3}
[ b(t), b^\dagger(t')]=\delta(t-t'), [ b(t), b(t')]=0.
\end{equation}

%, which can be obtained by calculating the evolution $b_{\rm in}(t)=\Theta^\dagger_t b_{\rm in}(0) \Theta_t$ with its initial state $b_{\rm in}(0)=\frac{1}{2\pi}\int_{-\infty}^{+\infty}b(\omega){\rm d}\omega$.
The field may be treated as a quantum stochastic process.
Note that we assume the initial state of the field on the Fock space $\mathfrak{F}$ is a vacuum state such that this process is analogous to Gaussian white noise with zero mean.

With the definition of the field (\ref{5}), we define an integrated operator by $B_t=\int_{t_0}^t b(t'){\rm d}t'$, $B_t^\dagger=\int_{t_0}^t b^\dagger(t'){\rm d}t'$
where $[B_t,B_{t'}^\dagger]={\rm min}(t,t'), [B_t,B_{t'}]=0$ and thus the operator $Q_t=B_t+B_t^\dagger$ is the quantum analog of the Wiener process and $q(t)=b(t)+b^\dagger(t)$ is quantum white noise. Hence, the evolution operator of the total system in the interaction picture $U_t=\Theta_t^\dagger \Lambda_t$
%can be expressed as
%\begin{equation}\label{9}
% \widetilde{U}_t=\Theta_t^\dagger U_t=\mathcal{T}{\rm exp}\int_0^t{\Big(}-\frac{{\rm i}}{\hbar}H_S{\rm d}t+L{\rm d}B_t^\dagger-L^\dagger{\rm d}B_t{\Big)}
%\end{equation}
%which
satisfies a quantum stochastic differential equation as follows
\begin{equation}\label{10}
  {\rm d}U_t={\big\{}-{\big(}{\rm i}H_S+\frac{1}{2}L^\dagger L{\big)}{\rm d}t+{\rm d}B_t^\dagger L-L^\dagger{\rm d}B_t{\big\}}U_t
\end{equation}
in It$\rm \bar{o}$ form.

By using (\ref{10}), a dynamical equation for an arbitrary operator $X$ of interest, namely a quantum Langevin equation, can be written down as
\begin{eqnarray}\label{11}
 {\rm d}X_t&=&{\big(}-{\rm i}[X_t,H_S(t)]+\mathcal{L}_{L_t}(X_t){\big)} {\rm d}t\nonumber\\
 &&+{\rm d}B_t^\dagger[X_t,L_t]+[L^\dagger_t,X_t]{\rm d}B_t
\end{eqnarray}
with a generator
\begin{equation}\label{11-1}
\mathcal{G}(X)=-{\rm i}[X,H_S]+\mathcal{L}_L(X),
\end{equation}
where $X_t=U^\dagger_t XU_t$, $L_t=U^\dagger_t LU_t$, and $H_S(t)=U^\dagger_tH_SU_t$. The notation $\mathcal{L}_{\cdot}(\cdot)$ defines a Lindblad superoperator which can be calculated as $\mathcal{L}_{N}(O)=\frac{1}{2}N^\dagger[O,N]+\frac{1}{2}[N^\dagger,O]N$ for two arbitrary operators $N$ and $O$ with suitable dimensions. Such an equation describes the dynamics of the system driven by an external white noise field, which has been widely used in the analysis and control of Markovian quantum systems~\cite{Gardiner}.

% and we have used $[\widetilde{U}_t,{\rm d}B_t]=[\widetilde{U}_t,{\rm d}B_t^\dagger]=0$ since $\widetilde{U}_t$ is an adapted process (see Gardiner, Quantum noise, pp.346).

\subsubsection{Input-output relations}\label{inputoutput}

To observe the dynamics of the system, one may consider an output field. The output field is the field after interaction with the system, which can be defined as $ B_{\rm out}(t)=U^\dagger_t B_t U_t$ satisfying a QSDE
\begin{equation}\label{13}
  {\rm d}B_{\rm out}(t)=L_t{\rm d}t+{\rm d}B_t,
\end{equation}
which shows the output field not only carries information of the system but also is affected by noise. As a result, the output field can be utilized by a quantum filter or a feedback controller~\cite{bouten,MY2003,ZGF2011}.
%with an system operator $L_t=\widetilde{U}_t^\dagger L\widetilde{U}_t$.
%See Refs. for more details.
\subsubsection{$(S,L,H)$ description}

In considering the interconnection of Markovian subsystems, an $(S,L,H)$ description has been developed.
A Markovian system $G$ can be systematically denoted as
\begin{equation}\label{13-11}
  G=(S,L,H),
\end{equation}
where the component $S$ is a scattering matrix describing the input-output relation of a field passing through beam splitters, the operator vector $L$ is a collection of system operators interacting with the external fields, and $H$ is the system Hamiltonian~\cite{Gough2009}.
%For example, considering a cavity system interacting with an external field, $G_a$ can be specified as
%\begin{equation}\label{13-12}
%  G_a=({\rm I},\sqrt{\gamma}a,\omega_0a^\dagger a).
%\end{equation}
%where $a$ and $a^\dagger$ are the annihilation and creation operators of the cavity modes defined on $\mathfrak{h}$. Note that the scattering matrix is set to be an identity matrix due to no interaction with beam splitters.

The $(S,L,H)$ description can also concisely describe the interconnection among subsystems.%, where the components $S,~L,~H$ of the total system can be directly calculated.
When we consider the connection between two subsystems, $G_1$ and $G_2$, they can be indirectly connected in a series product $G_1\lhd G_2$ or a concatenation product $G_1\boxplus G_2$ way via the input-output fields~\cite{Gough2009}. For a class of subsystems, they can also directly interact with each other which is denoted as $G_1\bowtie G_2$~\cite{ZGF2011}. With these basic interconnections, subsystems can be assembled into quantum feedback networks~\cite{Gough2009} or photonic networks~\cite{QPCM}.

\subsection{Ancillary system driven by white noise}
%\begin{figure}
%  % Requires \usepackage{graphicx}
%  \includegraphics[width=8.5cm]{f1.eps}\\
%  \caption{Ancillary system driven by white noise.}\label{f1}
%\end{figure}
To convert white noise to colored noise with a Lorentzian spectrum, we consider a Markovian linear quantum system, namely an ancillary system, whose mode can be taken as an internal mode of the non-Markovian environment. The ancillary system is described by $G_a=({\rm I},\sqrt{\gamma_0}a_0, \omega_0a^\dagger_0 a_0)$; i.e., an optical mode in a cavity, where $\omega_0$ is the angular frequency and $a_0$ ($a_0^\dagger$) is the annihilation (creation) operator of the ancillary system defined on a Hilbert space $\mathfrak{h}_0$. Here the coupling operator is chosen as $\sqrt{\gamma_0}a_0$, where $\sqrt{\gamma_0}$ is a damping rate with respect to the white noise field.

Let $U_t$ denote the unitary evolution for the ancillary system in the interaction picture with respect to the white noise field,
which satisfies a QSDE as follows
\begin{eqnarray}\label{16}
  {\rm d}U_t&=&{\big\{}-{\big(}{\rm i}\omega_0a^\dagger_0 a_0+\frac{\gamma_0}{2}a_0^\dagger a_0{\big)}{\rm d}t\nonumber\\
  &&~~~~~~~~~~~+\sqrt{\gamma_0}a_0{\rm d}B_t^\dagger-\sqrt{\gamma_0}a_0^\dagger{\rm d}B_t{\big\}}U_t.
\end{eqnarray}

The generator for the ancillary system is $\mathcal{G}_a(X_a)=-{\rm i}[X_a,\omega_0a^\dagger_0 a_0]+\mathcal{L}_{\sqrt{\gamma_0}a_0}(X_a)$,
where $X_a$ is an operator of the ancillary system. Note that since the ancillary system is driven by white noise, we continue to use the notation $U_t$ and ${\rm d}B_t$ to describe the system evolution operator and white noise process, respectively. Hence, a QSDE for the operator $a_0$ can be obtained as
\begin{equation}\label{14}
 {\rm d}a_0(t)=-(\frac{\gamma_0}{2}+{\rm i}\omega_0)a_0(t){\rm d}t-\sqrt{\gamma_0}{\rm d}B_t
\end{equation}
with $a_0(t)=U_t^\dagger a_0U_t$.

We define
\begin{equation}\label{14-1}
 c(t)=-\frac{\sqrt{\gamma_0}}{2}a_0(t)
\end{equation}
as a fictitious output, which is different from the output field discussed in \ref{inputoutput}.
Then we have that the fictitious output $c(t)$ satisfies a QSDE as follows
\begin{equation}\label{15}
  {\rm d}c(t)=-(\frac{\gamma_0}{2}+{\rm i}\omega_0)c(t){\rm d}t+\frac{\gamma_0}{2}{\rm d}B_t
\end{equation}
whose formal solution can be expressed as
\begin{equation}\label{15-1}
c(t)=e^{-(\frac{\gamma_0}{2}+{\rm i}\omega_0)(t-t_0)}c(t_0)+\int_{t_0}^t\frac{\gamma_0}{2} e^{-(\frac{\gamma_0}{2}+{\rm i}\omega_0)(t-\tau)}{\rm d}B_\tau
\end{equation}
with an initial state $c(t_0)$.

\subsection{Lorentzian spectrum and broadband limit}

As a part of the environment, the dynamics of the ancillary system may be assumed to start from a long time ago in which case  we let $t_0\rightarrow -\infty$.
Hence we have
\begin{equation}\label{17}
  c(t)=U_t^\dagger cU_t=\int_{-\infty}^t\xi(t-\tau)b(\tau){\rm d}\tau
\end{equation}
 in place of (\ref{15-1}), which is a convolution involving the white noise field and the kernel
  $\xi(t)=\frac{\gamma_0}{2} e^{-(\frac{\gamma_0}{2}+{\rm i}\omega_0)t}$.
 %In the following, the symbol $c(t)$ refers to the stationary solution $c(t)$ as in (\ref{17}).
The power spectral density for  $c(t)$ is Lorentzian calculated to be
\begin{equation}\label{19}
  S(\omega)=\frac{\frac{\gamma_0^2}{4}}{\frac{\gamma_0^2}{4}+(\omega-\omega_0)^2}
\end{equation}
with a center frequency $\omega_0$ and a linewidth $\gamma_0$ determined by the angular frequency of the ancillary system and the damping rate with respect to the white noise field, respectively.

In this case, the commutation relation for $c(t)$ is calculated to be
\begin{equation}\label{20}
  [c(t),c^\dagger(t')]=\mathcal{M}(t-t')
\end{equation}
with the memory kernel function
\begin{equation}\label{21}
 \mathcal{M}(t-t')=\mathcal{F}^{-1}[S(\omega)]=\int_{-\infty}^t\xi(t-\tau)\xi^*(t'-\tau){\rm d}\tau,
\end{equation}
which is different from the delta commutation relation for the white noise (\ref{3}). Here, $\mathcal{F}^{-1}$ denotes the inverse Fourier transform.

This model has a finite bandwidth which is determined by $\gamma_0$.
In the broadband limit; i.e., $\gamma_0\rightarrow\infty$, we have
\begin{equation}\label{18}
  c(t) = \int_{-\infty}^t \frac{\gamma_0}{2} e^{-(\frac{\gamma_0}{2}+{\rm i}\omega_0)(t-\tau)} b(\tau){\rm d}\tau\approx b(t)
\end{equation}
where the fictitious output $c(t)$ reduces to white noise with delta correlation function. This means the dynamics of the ancillary system can be ignored in the broadband limit.

\section{Principal system interacting with the ancillary system}\label{sec3}
\subsection{Dynamics of the augmented system}
\begin{figure}
  \includegraphics[width=8.5cm]{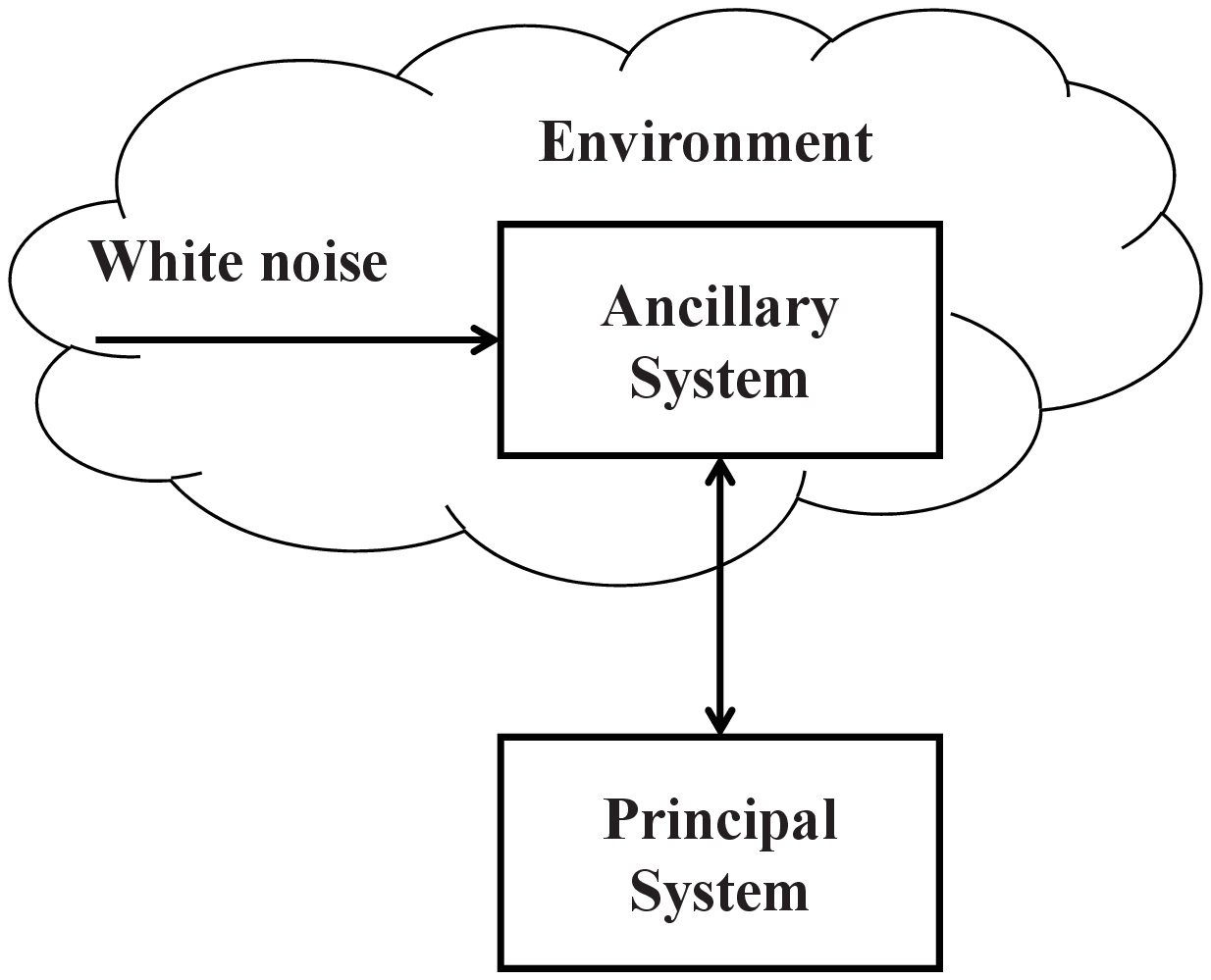}\\
  \caption{Schematic diagram for the direct coupling between the ancillary and the principal system.}\label{AP}
\end{figure}

A principal system $G_p$ on a Hilbert space $\mathfrak{h}$ with a free Hamiltonian $H_S$ is of interest, which is disturbed by the colored noise created by the noise model, i.e.,the ancillary system, via a direct interaction. Thus the principal system and the ancillary system constitute an augmented system. We assume that  the interaction Hamiltonian for the coupling between the principal system and the ancillary system is
\begin{equation}\label{22}
  H_I={\rm i}(c^\dagger Z-Z^\dagger c),~~~~Z=\sqrt{\kappa}K,
\end{equation}
where $Z$ is an direct coupling operator of the principal system expressed as
a product between a principal system operator $K$ and a coupling strength $\sqrt{\kappa}$. The principal and ancillary systems influence each other due to their direct interaction as shown in Fig.~\ref{AP}.
This augmented principal-ancillary system can be described by using an $(S,L,H)$ description as
\begin{equation}\label{23}
%&=&(-,-,H_S+H_I)\boxplus({\rm I},\sqrt{\gamma}a_0,\omega_0a^\dagger_0a_0)\nonumber\\
 G_{p,a}=({\rm I},\sqrt{\gamma_0}a_0,H_S+H_I+\omega_0a^\dagger_0a_0),
\end{equation}
where the evolution operator $\bar{U}_t$ of the total system satisfies a QSDE as follows
\begin{eqnarray}\label{23-2}
  {\rm d}{\bar U}_t&=&{\big\{}-{\rm i}{\big(}H_S+H_I+\omega_0a^\dagger_0 a_0{\big)}{\rm d}t-\frac{\gamma_0}{2}a_0^\dagger a_0{\rm d}t+\nonumber\\
  &&\sqrt{\gamma_0}{\rm d}B_t^\dagger a_0-\sqrt{\gamma_0}a_0^\dagger{\rm d}B_t{\big\}}{\bar U}_t.
\end{eqnarray}

Let $ X'$ denote any operator for the augmented principal and ancillary system. Its evolution can be defined as $\bar {X}'_t=\bar U_t^\dagger  X'\bar U_t$ which satisfies a QSDE written as
\begin{eqnarray}\label{38}
  {\rm d}\bar {X}'_t&=&-{\rm i}[\bar {X}'_t,\bar {H}_t]{\rm d}t+\mathcal{L}_{\sqrt{\gamma_0}\bar {a}_{0}(t)}(\bar {X}'_t){\rm d}t\nonumber\\
  &&+([\bar {X}'_t,\bar {c}^\dagger_t \bar {Z}_t]+[\bar {Z}_t^\dagger\bar {c}_t,\bar {X}'_t]){\rm d}t\nonumber\\
  %\frac{\gamma_0}{2}(\bar {a}_{0,t}^\dagger[\bar {X}'_t,\bar {a}_{0,t}]+[\bar {a}_{0,t}^\dagger,\bar {X}'_t] \bar {a}_{0,t}){\rm d}t\nonumber\\
  &&+\sqrt{\gamma_0}({\rm d}B_t^\dagger[\bar {X}'_t,\bar {a}_{0}(t)]+[\bar {a}_{0}^\dagger(t),\bar {X}'_t]{\rm d}B_t),
\end{eqnarray}
with $\bar {H}_t=\bar U_t^\dagger(H_S+\omega_0a_0^\dagger a_0)\bar U_t$, $\bar {a}_{0}(t)=\bar U_t^\dagger a_0 \bar U_t$, $\bar {c}_t=\bar U_t^\dagger c \bar U_t$, and $\bar {Z}_t=\bar U_t^\dagger Z \bar U_t$.

In particular, for $X'=X $ a principal system operator, Eq.~(\ref{38}) reduces to
\begin{equation}\label{39}
  {\rm d}\bar{X}_t=-{\rm i}[\bar{X}_t,\bar{H}_S(t)]{\rm d}t+(\bar{c}^\dagger_t[\bar{X}_t, \bar{Z}_t]+[\bar{Z}_t^\dagger,\bar{X}_t]\bar{c}_t){\rm d}t,
\end{equation}
with $\bar{X}_t=\bar U_t^\dagger X \bar U_t$ and $\bar {H}_S(t)=\bar U_t^\dagger H_S\bar U_t$.
When $X'=c=-\frac{\sqrt{\gamma_0}}{2}a_0$, i.e., for an operator of the ancillary system, we have
\begin{equation}\label{39-1}
{\rm d}\bar{c}(t)=-(\frac{\gamma_0}{2}+{\rm i}\omega_0)\bar{c}(t){\rm d}t+\frac{\gamma_0}{4}\bar{Z}_t{\rm d}t+\frac{\gamma_0}{2}{\rm d}B_t.
\end{equation}
Note that  $\bar{c}(t)$ can be written as
\begin{equation}\label{39-2}
\bar{c}(t)=c(t)+\frac{1}{2}\int_{t_0}^t\xi(t-\tau)\bar{Z}_\tau{\rm d}\tau,
\end{equation}
which shows the ancillary system not only depends on the Lorentzian noise $c(t)$ defined in (\ref{15-1}) but also is disturbed by the principal system as indicated by the integral term  in (\ref{39-2}).
 %Although $\bar{c}(t)\neq c(t)$, we may witness the system disturbed by $c(t)$ with a Lorenztian spectrum in an interaction picture.
\subsection{Interaction picture with respect to the ancillary system}
We can move to an interaction picture with respect to the ancillary system by defining an evolution operator as $V_t=U_t^\dagger\bar{U}_t$, whose evolution satisfies
\begin{equation}\label{27}
\dot V_t={\big\{}-{\rm i}H_S-(Z^\dagger c(t)-c^\dagger(t)Z){\big\}}{V}_t.
\end{equation}
In this interaction picture, the system is described by
\begin{equation}\label{28}
  G=(-,-,H_S+{\rm i}(c^\dagger(t)Z-Z^\dagger c(t)))
\end{equation}
where $c(t)$ is given by (\ref{14}) with a Lorentzian spectrum.

Note that in the interaction picture the system is driven by Lorentzian noise as given in Eqs.(\ref{27}) and (\ref{28}). The evolution of an operator $X$ for the principal system in the interaction picture is equivalent to that in the augmented system due to
\begin{equation}\label{28-1}
  V_t^\dagger X V_t=\bar{U}_t^\dagger U_t X U_t^\dagger\bar{U}_t=\bar{U}_t^\dagger XU_t U_t^\dagger\bar{U}_t=\bar{U}_t^\dagger X\bar{U}_t.
\end{equation}
Hence, the operator evolution for the principal system in Eq.~(\ref{39}) is disturbed by Lorentzian noise as well.
\subsection{Non-Markovian dynamics of the principal system and its Markovian limit}
Substituting the solution (\ref{39-2}) into (\ref{39}), a non-Markovian Langevin equation for the principal system can be obtained as
\begin{eqnarray}\label{32-1}
   \dot{\bar{X}}_t &=& -{\rm i}[\bar{X}_t,\bar{H}_S(t)]+c^\dagger(t)[\bar{X}_t,\bar{Z}_t]+[\bar{Z}_t^\dagger,\bar{X}_t]c(t)\nonumber\\
   &&+D(\xi^*,\bar{Z}^\dagger)_t[\bar{X}_t,\bar{Z}_t]+[\bar{Z}_t ^\dagger,\bar{X}_t]D(\xi,\bar{Z})_t
\end{eqnarray}
where the convolution terms are expressed as
\begin{eqnarray}\label{32-2}
  D(\xi,\bar{Z})_t&=&\frac{1}{2}\int_{t_0}^t \xi(t-\tau)\bar{Z}_\tau{\rm d}\tau.
 % D(\xi,b)_t&=&\int_{-\infty}^t \xi(t-\tau){\rm d}B_{\tau}.
\end{eqnarray}
This Langevin equation coincides with the existing non-Markovian Langevin equations whose integral terms represent the memory effect~\cite{XuePRA2012,Tan2011}.

\textit{Remark}: Note that taking the broadband limit $\gamma_0\rightarrow+\infty$, i.e., using the quantum Wong-Zakai theorem~\cite{Gough06JMP},  the Langevin equation (\ref{32-1}) reduces to
\begin{eqnarray}\label{32-4}
   \dot{\bar{X}}_t &=& -{\rm i}[\bar{X}_t,\bar{H}_S(t)]+\mathcal{L}_{\sqrt{\kappa}\bar{K}_t}(\bar{X}_t)\nonumber\\
   &&+\sqrt{\kappa}b^\dagger(t)[\bar{X}_t,\bar{K}_t]+\sqrt{\kappa}[\bar{K}_t^\dagger,\bar{X}_t]b(t),
\end{eqnarray}
with $\bar{K}_t=\bar{U}^\dagger_tK\bar{U}_t$, which is coincident with the Markovian Langevin equation~\cite{Gardiner}.
\subsection{Master equation}

By using the fact that the expectation of an operator $X$ in the Heisenberg picture is equal to that in the Schr${\rm \ddot{o}}$dinger picture, we can obtain an unconditional master equation for the augmented principal and ancillary system as
\begin{eqnarray}\label{40}
 \dot \rho_t&=&-{\rm i}[H_S+\omega_0a_0^\dagger a_0,\rho_t]+\mathcal{L}^*_{\sqrt{\gamma_0}a_0}(\rho_t)\nonumber\\
 &&+[c^\dagger Z,\rho_t]+[\rho_t, Z^\dagger c],%\frac{\gamma_0}{2}([a_0\rho^{pa}_t,a_0^\dagger]+[a_0,\rho^{pa}_t a_0^\dagger]).
\end{eqnarray}
where $\rho_t$ is the unconditional state of the augmented system and the superoperator $\mathcal{L}^*_\cdot(\cdot)$ is the adjoint of the Lindblad superoperator calculated as $\mathcal{L}^*_N(O)=\frac{1}{2}N[O,N^\dagger]+\frac{1}{2}[N,O]N^\dagger$ for arbitrary operators $N$ and $O$ with suitable dimensions.

As shown in Eq.~(\ref{40}), the state evolution of the augmented principal and ancillary system is Markovian, where the state variation only depends on the present state. One can also obtain the unconditional state $\rho^p_t$ of the principal system by calculating
\begin{equation}\label{41}
  \rho^p_t={\rm tr}_a[\rho_t],
\end{equation}
which will not satisfy a Markovian evolution. Note that ${\rm tr}_a[\cdot]$ means the partial trace with respect to the ancillary system.

\section{Quantum filtering for non-Markovian quantum systems}\label{sec4}
\subsection{The augmented system under a probing field}

\begin{figure}
  \includegraphics[width=8.5cm]{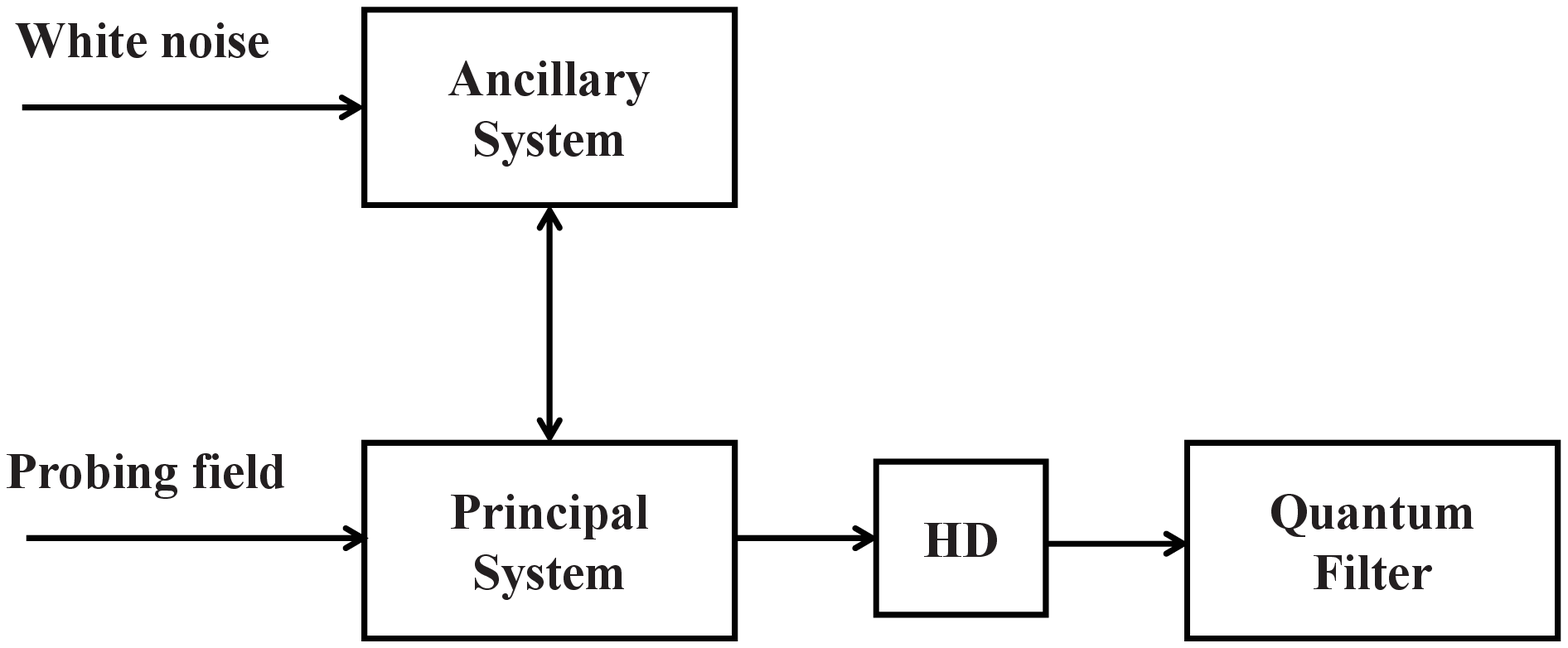}\\
  \caption{Schematic diagram for probing a non-Markovian quantum system.}\label{Filtering}
\end{figure}

To estimate the dynamics of the non-Markovian system, we can construct a quantum filter using a probing field defined on a Fock space $\mathfrak{F}_1$ as shown in Fig.~\ref{Filtering}. The total system $G_{T}$ can be described as
\begin{eqnarray}\label{47}
  G_{T}&=&({\rm I},\left(
                     \begin{array}{c}
                      \sqrt{\gamma_0}a_0 \\
                       {L} \\
                     \end{array}
                   \right)
,H_S+H_I+\omega_0a^\dagger_0a_0)
\end{eqnarray}
where $L$ is the coupling operator of the principal system for the probing field. We denote the evolution operator of the total system as $\tilde{U}_t$
which satisfies a QSDE as follows
\begin{eqnarray}\label{49}
  {\rm d}\tilde{U}_t&=&{\big\{}-{\rm i}{\big(}H_S+H_I+\omega_0a^\dagger_0 a_0{\big)}{\rm d}t-\frac{\gamma_0}{2}a_0^\dagger a_0{\rm d}t\nonumber\\
  &&-\frac{1}{2}{L}^\dagger {L}{\rm d}t+\sqrt{\gamma_0}{\rm d}B_t^\dagger a_0-\sqrt{\gamma_0}a_0^\dagger{\rm d}B_t\nonumber\\
  &&+{\rm d}{\tilde B}_t^{\dagger}{L}-{L}^{\dagger}{\rm d}{\tilde B}_t{\big\}}\tilde{U}_t.
\end{eqnarray}
Then a QSDE for an operator $ X'$ of the augmented principal and ancillary systems defined on $\mathfrak{h}\otimes\mathfrak{h}_0$ can be derived as
\begin{eqnarray}\label{49-1}
  {\rm d}\tilde{X}'_t&=&-{\rm i}[\tilde{X}'_t,\tilde{H}_t]{\rm d}t+([\tilde{X}'_t,\tilde{c}^\dagger_t \tilde{Z}_t]+[\tilde{Z}_t^\dagger\tilde{c}_t,\tilde{X}'_t]){\rm d}t\nonumber\\
  &&+(\mathcal{L}_{\sqrt{\gamma_0}\tilde{a}_{0}(t)}(\tilde{X}'_t)+\mathcal{L}_{\tilde L_t}(\tilde{X}'_t)){\rm d}t\nonumber\\%\frac{\gamma_0}{2}(\tilde{a}_{0,t}^\dagger[\tilde{X}'_t,\tilde{a}_{0,t}]+[\tilde{a}_{0,t}^\dagger,\tilde{X}'_t] \tilde{a}_{0,t})
  &&+\sqrt{\gamma_0}({\rm d}B_t^\dagger[\tilde{X}'_t,\tilde{a}_{0}(t)]+[\tilde{a}_{0}^\dagger(t),\tilde{X}'_t]{\rm d}B_t)\nonumber\\
  %&&+\frac{1}{2}(\tilde {L}^\dagger_t[\tilde X'_t,\tilde L_t]+[\tilde {L}_t^\dagger,\tilde X'_t]\tilde L_t){\rm d}t\nonumber\\
   &&+{\rm d}{\tilde B}_t^{\dagger}[\tilde X'_t,\tilde L_t]+[\tilde {L}^\dagger_t,\tilde X'_t]{\rm d}{\tilde B}_t,
\end{eqnarray}
where $\tilde {X}'_t=\tilde{U}_t^\dagger X' \tilde{U}_t$, $\tilde H_t=\tilde{U}_t^\dagger (H_S+\omega_0 a^\dagger_0 a_0)\tilde{U}_t$, $\tilde{c}_t=\tilde{U}_t^\dagger c \tilde{U}_t$, $\tilde{Z}_t=\tilde{U}_t^\dagger Z \tilde{U}_t$, $\tilde{L}_t=\tilde{U}_t^\dagger L \tilde{U}_t$, $\tilde{a}_{0}(t)=\tilde{U}_t^\dagger a_{0}\tilde{U}_t$ and ${\rm d}{\tilde B}_t$ is the probing field process.

Note that supposing an operator of the augmented system can be denoted as $X'=X_p\otimes X_a$, the generator can be written as
\begin{eqnarray}
  \mathcal{G}_T(X') &=& \mathcal{G}_p(X_p)\otimes X_a+X_p\otimes  \mathcal{G}_a(X_a) \nonumber \\
&& -{\rm i}[X',H_I],
\end{eqnarray}
where $\mathcal{G}_p(X_p)=-{\rm i}[X_p,H_S]+\mathcal{L}_L(X_p)$ and $\mathcal{G}_a(X_a)=-{\rm i}[X_a,\omega_0 a_0^\dagger a_0]+\mathcal{L}_{\sqrt{\gamma_0}a_0}(X_a)$
are the generators for the principal system and the ancillary system, respectively.

\subsection{Belavkin quantum filter}

Using the probing field, the system can be continuously monitored via homodyne detection, where a quadrature of the probing field is detected and can be used as input to a quantum filter.

The output $Y_t$ satisfying the QSDE ${\rm d}Y_t=(\tilde L_t+\tilde {L}^\dagger_t){\rm d}t+{\rm d}Q_t$
with $Q_t=\tilde B_t+\tilde B_t^\dagger$, commutes with an observable of the augmented system, i.e., a non-demolition condition
\begin{equation}\label{61}
 [\tilde X'_t, Y_\tau]=0,~0\leq\tau\leq t,
\end{equation}
is satisfied which means the continuous measurement of the field does not change the observable of the system. Hence, a quantum filter using the output process $\{Y_\tau, 0\leq \tau\leq t\}$ for the estimation of the evolution of the system observable $\tilde X'_t$ can be constructed based on. Note that we have assumed that the detection efficiency of the homodyne detector is perfect, i.e., $100\%$ efficiency, and the state of the field is in a vacuum state.

%In the Heisenberg picture, the expected value for an operator with a given state $\rho_0\otimes\Phi$ can be defined as
%\begin{equation}\label{62}
%  \mathbb{E}[\cdot]={\rm tr}[(\rho_0\otimes\Phi)(\cdot)],
%\end{equation}
%where $\rho_0$ and $\Phi$ are the initial state of the augmented system and the probing field, respectively.
The estimate of an observable $\tilde X_t'$ is defined by a conditional expectation as
\begin{equation}\label{63}
  \hat X'_t=\pi_t( X')=\mathbb{E}[\tilde X'_t|\mathcal{Y}_t],
\end{equation}
where $\mathcal{Y}_t$ is a commutative subspace of operators generated by the measurement results $Y(\tau),~0\leq\tau\leq t$. The conditional expectation can be interpreted as the orthogonal projection of $\tilde X'_t$ onto a subspace of $\mathcal{Y}_t$, which means that $\tilde X'_t-\pi_t(X')$ is orthogonal to the measurement subspace $\mathcal{Y}_t$, i.e.,
\begin{equation}\label{64}
    \mathbb{E}[(\pi_t(X')-\tilde X'_t)C_t]=0
\end{equation}
for an arbitrary operator $C_t$ in $\mathcal{Y}_t$~\cite{bouten,belavkin}.

Hence, we can obtain a \textit{Belavkin} quantum filter for the augmented system as
\begin{eqnarray}\label{67}
    {\rm d}\pi_t( X')&=& \pi_t(\mathcal{G}_T (X')){\rm d}t-(\pi_t( X' L+{ L}^\dagger  X')-\pi_t( X')\nonumber\\
    &&\times\pi_t( L+ {L}^\dagger))({\rm d}Y_t-\pi_t( L+ {L}^\dagger)  {\rm d}t)
\end{eqnarray}
where ${\rm d}W={\rm d}Y_t-\pi_t( L+ {L}^\dagger)  {\rm d}t$ and $W$ is called the innovation process and is equivalent to a classical Wiener process. Note that the increment ${\rm d}W$ is independent of $\pi_\tau( X'), 0\leq\tau\leq t$.
\subsection{Stochastic master equation}

The conditional expectation $\pi_t( X')$ is defined for the augmented system and thus a conditional density matrix $\hat \rho_t$ for the augmented system can be defined by
\begin{equation}\label{68}
  \pi_t( X')={\rm tr}[\hat \rho_t X'].
\end{equation}
Hence, a stochastic master equation for the augmented system can be obtained from the quantum filter (\ref{67}) as
\begin{eqnarray}\label{69}
 {\rm d}\hat \rho_t &=&\mathcal{G}_T^*(\hat \rho_t){\rm d}t+\mathcal{F}_{ L}(\hat \rho_t){\rm d}W
%  &=&(-{\rm i}[H_S+\omega_0a_0^\dagger a_0,\hat \rho_t]+\mathcal{L}^*_{ L}(\hat \rho_t)+ \mathcal{L}^*_{\sqrt{\gamma_0} a_0}(\hat  \rho_t)\nonumber\\
% &&+[\hat \rho_t,Z^\dagger c]+[c^\dagger Z,\hat \rho_t]){\rm d}t+\mathcal{F}_{ L}(\hat \rho_t){\rm d}W
\end{eqnarray}
with $\mathcal{F}_{ L}(\hat \rho_t)= L\hat \rho_t+\hat \rho_t {L}^\dagger-{\rm tr}[( L+ {L}^\dagger)\hat \rho_t]\hat \rho_t$,
which is a Markovian stochastic master equation. The superoperator $\mathcal{G}_T^*$ is the adjoint of $\mathcal{G}_T$.

However, a stochastic master equation for the density operator $\hat \rho_t^p$ of the principal system of interest is not in a Markovian form. Instead, we can trace out the ancillary system to obtain the conditional state of the principal system $\hat \rho_t^p$  as
\begin{equation}\label{72}
  \hat \rho_t^p={\rm tr}_a[\hat \rho_t].
\end{equation}
In practice, one cannot obtain an exact description of $\hat \rho_t^p$ due to the infinite dimensional nature of the ancillary system. While a truncation can be made for the ancillary system, i.e, we can assume it is a $N$-level system and thus it is possible to calculate an approximation to the partial trace (\ref{72}).

\section{An illustrative example}\label{sec5}
%\subsubsection{Spectrum of the output for the probing field}
%\begin{figure}
%  % Requires \usepackage{graphicx}
%  \includegraphics[width=8.5cm]{output.eps}\\
%  \caption{Schematic plot for observing the spectrum of the output for the probing field.}\label{output}
%\end{figure}
\begin{figure}
  \includegraphics[width=8.5cm]{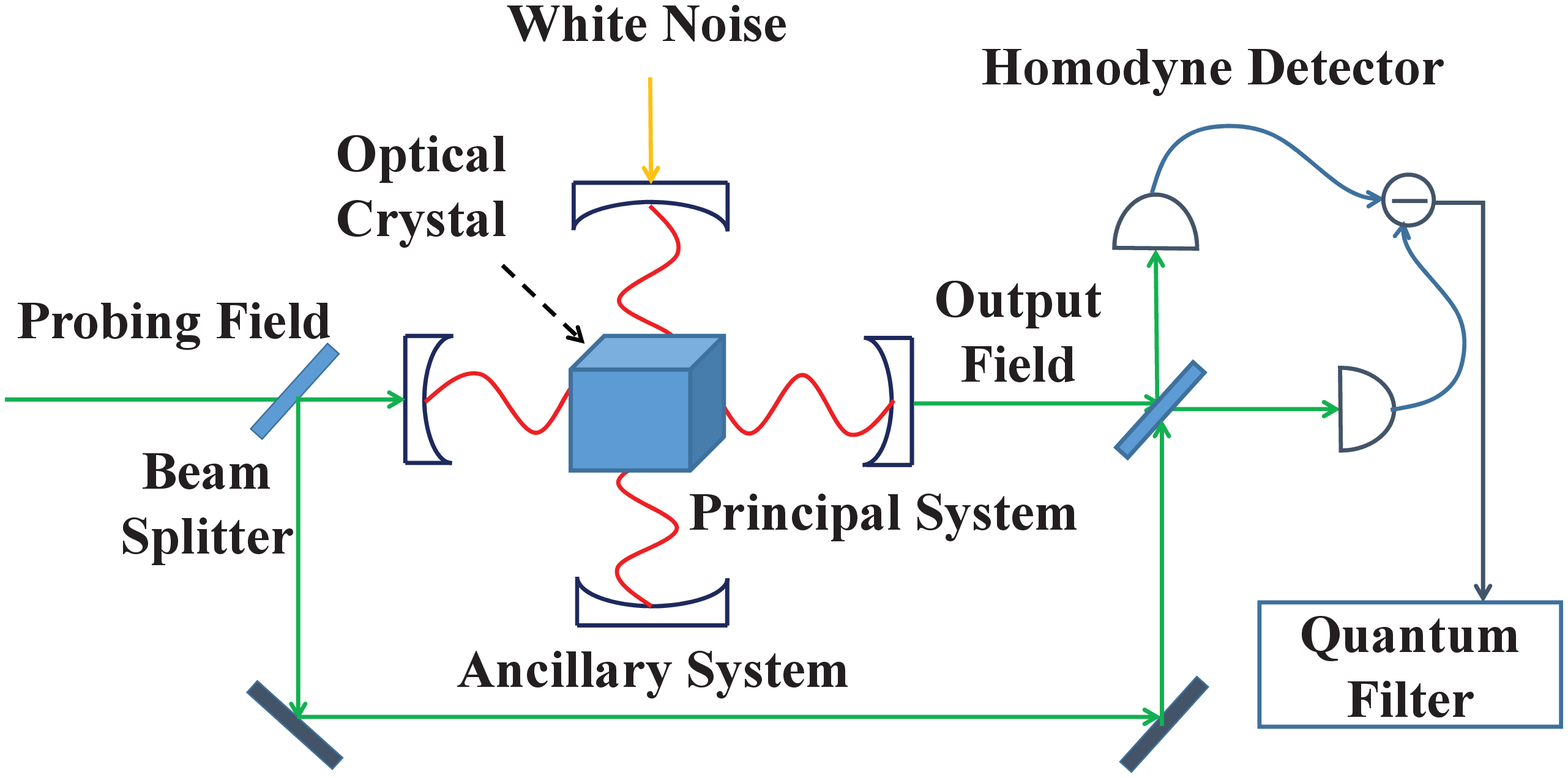}\\
  \caption{An illustrative example of optical systems.}\label{example-opc}
\end{figure}

In this section, an optical realization of a non-Markovian quantum system is shown in Fig.~\ref{example-opc}. The ancillary cavity, which is vertically oriented, is pumped by white noise. Another cavity is orthogonally oriented to the ancillary cavity and can be considered as the principal system. The optical modes in the two cavities are directly and strongly coupled by an optical crystal. A probing field  is applied to the principal cavity, whose output is observed via homodyne detection in order to apply a quantum filter to the system.

The Hamiltonian of the principal system can be written as $H_S=\omega_s a_s^\dagger a_s$ with an angular frequency $\omega_s$ and an annihilation (creation) operator $a_s$ ($a_s^\dagger$). Then, the coupling operator $Z$ and $L$ can be specified as $Z=\sqrt{\kappa}a_s$ and $L=\sqrt{\gamma_1}a_s$. Substituting these operators into Eq.~(\ref{49-1}), Langevin equations for the augmented principal and ancillary cavities can be written as
\begin{eqnarray}\label{72-1}
\left[
        \begin{array}{c}
          \dot{a}_s(t)\\
         \dot{ a}_0(t)\\
        \end{array}
      \right]&=&\left[
                \begin{array}{cc}
                  -{\rm i}\omega_s-\frac{\gamma_1}{2} & \frac{\sqrt{\kappa\gamma_0}}{2} \\
                  -\frac{\sqrt{\kappa\gamma_0}}{2}  & -{\rm i}\omega_0-\frac{\gamma_0}{2}\\
                \end{array}
              \right]\left[
        \begin{array}{c}
          a_s(t)\\
          a_0(t)\\
        \end{array}
      \right]\nonumber\\
      &&~~~~~~~~~~~~~~~~~~~~~~~-\left[
        \begin{array}{c}
          \sqrt{\gamma_1}b_p(t)\\
          \sqrt{\gamma_0}b(t)\\
        \end{array}
      \right],
\end{eqnarray}
where $b_p(t)$ and $b(t)$ are the probing field and white noise field, respectively.
The output equation with respect to the probing field $b_p(t)$ is
\begin{equation}\label{72-2}
  b_{\rm out}(t)=b_p(t)+\sqrt{\gamma_1}a_s(t).
\end{equation}

Since the operators in Eq.~(\ref{72-1}) are not self-adjoint operators and the coefficients may be complex valued, it is convenient to express Eq.~(\ref{72-1}) and~(\ref{72-2}) as
\begin{eqnarray}\label{72-2-1}
  \dot{ x}(t)
   &=& A { x}(t)+B u(t)\\
 y(t)
   &=&C\left[
        \begin{array}{cc}
          I & 0 \\
        \end{array}
      \right]{ x}(t)+\left[
        \begin{array}{cc}
          I& 0 \\
        \end{array}
      \right] u(t)
\end{eqnarray}
with\begin{eqnarray}
      A &=& \left[
              \begin{array}{cccc}
                -\frac{\gamma_1}{2} & \omega_s & \frac{\sqrt{\kappa\gamma_0}}{2}  & 0 \\
                -\omega_s& -\frac{\gamma_1}{2} & 0 &\frac{ \sqrt{\kappa\gamma_0}}{2} \\
                -\frac{\sqrt{\kappa\gamma_0}}{2}& 0 & -\frac{\gamma_0}{2} & \omega_0 \\
                0 & -\frac{\sqrt{\kappa\gamma_0}}{2} & -\omega_0 & -\frac{\gamma_0}{2} \\
              \end{array}
            \right],\\
     B &=& {\rm diag}[-\sqrt{\gamma_1},-\sqrt{\gamma_1},-\sqrt{\gamma_0},-\sqrt{\gamma_0}], \\
     C&=& {\rm diag}[\sqrt{\gamma_1},\sqrt{\gamma_1}]
    \end{eqnarray}
    in a quadrature representation with self adjoint operators and real valued coefficients,
where $ x(t)=[ q_s(t),p_s(t), q_0(t),p_0(t)]^T$, $ u(t)=[v_p(t),v_q(t),w_p(t),w_q(t)]^T$, and $  y(t)=[y_p(t),y_q(t)]^T$
are the quadrature representations of the operators of the principal and ancillary systems, the probing and white noise fields, and the outputs of the probing field, respectively.
By using a transformation matrix $\Xi=\frac{1}{\sqrt{2}}\left[
                          \begin{array}{cc}
                            1 & 1 \\
                            -{\rm i} &{\rm i} \\
                          \end{array}
                        \right]$, the components of $ x(t)$, $u(t)$, and $ y(t)$ are calculated as $[ q_s(t),p_s(t)]^T=\Xi[a_s(t),a_s^\dagger(t)]^T$, $[q_0(t),p_0(t)]^T=\Xi[a_0(t),a_0^\dagger(t)]^T$, $[v_p(t),v_q(t)]^T=\Xi[b_p(t),b_p^\dagger(t)]^T$, $[w_p(t),w_q(t)]^T=\Xi[b(t),b^\dagger(t)]^T$,$[y_p(t),y_q(t)]^T=\Xi[ b_{\rm out}(t), b_{\rm out}^\dagger(t)]^T$.

\subsection{The mean of the principal system}
Assume the unconditional state of the augmented system is Gaussian~\cite{olivares} and thus the means of operators for the system (\ref{72-2-1}) can be used to observe the system dynamics. Hence, we consider $ m(t)=\langle  x(t)\rangle$  satisfying
\begin{equation}\label{72-2-3}
\dot{ m}(t)=A m(t),
\end{equation}
where the quantum expectation $\langle\cdot\rangle={\rm tr}[\cdot \rho_0]$ is with respect to the initial state of the augmented system $\rho_0$.
%Also, a symmetrized covariance matrix $\hat V(t)$ defined as $\hat V(t)=\langle\frac{1}{2}[\Delta\hat x(t)\Delta\hat x^T(t)+(\Delta\hat x(t)\Delta\hat x^T(t))^T]\rangle$ with $\Delta\hat x(t)=\hat x(t)-\langle\hat x(t)\rangle$ is dominated by a Lyapunov matrix differential equation
%\begin{equation}\label{72-2-4}
%\dot{\hat V}(t)=A\hat V(t)+\hat V(t)A^T+BTB^T
%\end{equation}
%with the matrix $T=\frac{1}{2}\langle\hat u(t)\hat u^T(t)+(\hat u(t)\hat u^T(t))^T\rangle$. Note that if we partition the mean $\hat m(t)$ and the covariance matrix $\hat V(t)$ as $\hat m(t)=[\hat m_s(t),\hat m_0(t)]^T$ and $\hat V(t)=\left[
%                                                                                          \begin{array}{cc}
%                                                                                            \hat V_s(t) & \hat V_{s0}(t) \\
%                                                                                            \hat V_{s0}^T(t)& \hat V_0(t) \\
%                                                                                          \end{array}
%                                                                                        \right]$, respectively, the blocks $\hat m_s(t)$ and
%$\hat V_s(t)$ correspond to the mean and covariance of the principal system.

% which is characterized by the mean $\hat m(t)$ and covariance $\hat V(t)$. %Note that we ignore the results about the covariance $\hat V(t)$ due to no significant

%two cavities  are initially prepared in a separable state, i.e., $\hat V(0)=2I$, $T=\frac{1}{2}I$ and
On the other hand, the quantum filter (\ref{67}) as the estimation of the principal and ancillary systems is actually a quantum Kalman filter due to the linearity of the augmented system~\cite{PhysRevA.60.2700,PhysRevA.94.070405}. The conditional system dynamics are governed by the quantum Kalman filter as follows
\begin{eqnarray}
  {\rm d}\hat x_t &=& A\hat x_t  {\rm d}t+(\hat V_tF^T+\Sigma^T{\rm Im}(E)^T){\rm d}W\\
  \dot{\hat V}_t&=& A\hat V_t+\hat V_tA^T+D-(\hat V_tF^T+\Sigma^T{\rm Im}(E)^T)\nonumber\\
  &&\times(F\hat V_t+ {\rm Im}(E)\Sigma)
\end{eqnarray}
with $\Sigma=\bigoplus_{n=1}^2\left[
                                 \begin{array}{cc}
                                   0 & 1 \\
                                   -1 & 0\\
                                 \end{array}
                               \right]
$, $L=Ex=[\sqrt{\gamma_1}\Xi^{-1},\textbf{0}]x$, $F=E+E^*$, and $D=\Sigma{\rm Re}(E^\dagger E)\Sigma^T$, where $\hat x(t)$ is the conditional expectation of $x(t)$ and $\hat V_t$ is a symmetrized covariance matrix. The bold $\textbf{0}$ is a $2\times2$ zero matrix.

\begin{figure}
  \includegraphics[width=8.5cm]{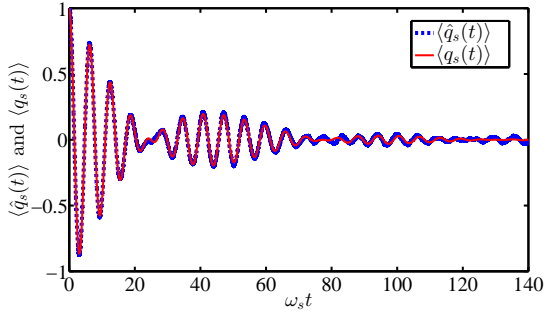}\\
  \caption{The dynamics of the unconditional and conditional means of the position components for the principal system with $\kappa=0.6$, $\Delta=0, \gamma_0=0.6$, and $\gamma_1=0.8$.}\label{filtering4}
\end{figure}

For simplicity, we choose the parameters of the system as $\omega_s=\omega_0=10{\rm GHz}$, $\kappa=0.6$, $\gamma_0=0.6$, and $\gamma_1=0.8$ and assume that the initial mean $ m(0)=[1,0,0,0]^T$ of the unconditional state is the same as the initial conditional state expectation of the Kalman filter, i.e., $\hat x(0)=[1,0,0,0]^T$ where the first element of $\hat x(0)$ is $\langle\hat q_s(0)\rangle=1$. In Fig.~\ref{filtering4}, the red solid line is the trajectory of the mean of the unconditional state $q_s(t)$ for the principal system. The oscillations of the curve envelopes are caused by the disturbance of the ancillary system, which indicates that energy is exchanged between the principal and the ancillary system showing non-Markovian characteristics. Compared with the unconditional trajectory, the average trajectory of the conditional expectation of position for 1000 realizations denoted as $\langle \hat q_s(t)\rangle$, is plotted as the blue dashed line in Fig.~\ref{filtering4}. This shows that the blue line for the quantum filter can match with the red line for the unconditional state of the system, i.e., the quantum filter can estimate the unconditional state of the system.

%\begin{figure}
%\subfigure[The dynamics of $\langle q_s(t)\rangle$]{
%\begin{minipage}[b]{.5\textwidth}
%\includegraphics[width=1\textwidth]{meanQst.eps}\label{meanQst}
%\end{minipage}
%}\\
%\subfigure[The dynamics of $\langle p_s(t)\rangle$]{
%\begin{minipage}[b]{.5\textwidth}
%\includegraphics[width=1\textwidth]{meanPst.eps}\label{meanPst}
%\end{minipage}
%}
%  \caption{.}\label{mean}
%\end{figure}
%\begin{figure}
%\begin{minipage}[b]{.5\textwidth}
%\includegraphics[width=1\textwidth]{v11.eps}\label{v11}
%\end{minipage}
%  \caption{.}\label{v11}
%\end{figure}

%\begin{figure}
%\begin{minipage}[b]{.5\textwidth}
%\includegraphics[width=1\textwidth]{Qspec.eps}\label{Qspec}
%\end{minipage}
%  \caption{.}\label{Qspec}
%\end{figure}
\subsection{The spectrum with respect to the white noise field}
Alternatively, it is useful to observe the output field spectrum as considering the properties of the system~\cite{shabani2014}.
To calculate the power spectral density of the output of the probing field, we can transform (\ref{72-1}) and (\ref{72-2}) into the frequency domain via the Fourier transform and express the output in terms of the white noise field and the probing field as
\begin{equation}\label{72-3}
  b_{\rm out}({\rm i}\omega)=G_1({\rm i}\omega)b_p({\rm i}\omega)+G_2({\rm i}\omega)b({\rm i}\omega)
\end{equation}
in the frequency domain, where $b_{\rm out}({\rm i}\omega)$, $b_p({\rm i}\omega)$ and $b({\rm i}\omega)$ are the Fourier transform of $b_{\rm out}(t)$, $b_p(t)$ and $b(t)$, respectively. The transfer function with respect to the probing field and the white noise field can be calculated as
\begin{equation}
       G_1({\rm i}\omega) =\frac{({\rm i}(\omega_s-\omega)-\frac{\gamma_1}{2})({\rm i}(\omega_0-\omega)+\frac{\gamma_0}{2})+\kappa\gamma_0}{({\rm i}(\omega_s-\omega)+\frac{\gamma_1}{2})({\rm i}(\omega_0-\omega)+\frac{\gamma_0}{2})+\kappa\gamma_0},
      \end{equation}
 \begin{equation}
       G_2({\rm i}\omega) = \frac{-\gamma_0\sqrt{\kappa\gamma_1}}{({\rm i}(\omega_s-\omega)+\frac{\gamma_1}{2})({\rm i}(\omega_0-\omega)+\frac{\gamma_0}{2})+\kappa\gamma_0},
     \end{equation}
respectively.

By detecting the quadrature of the output field $X_{\rm out}({\rm i}\omega)=\frac{1}{2}[ b_{\rm out}({\rm i}\omega)+ b_{\rm out}^\dagger({\rm i}\omega)]$, the power spectral density of the output field can be calculated as
\begin{equation}\label{72-5}
  S(\tilde \omega):=\langle|X_{\rm out}(\tilde \omega)|^2\rangle=\frac{1}{4}(|G_1(\tilde \omega)|^2+|G_2(\Tilde \omega)|^2)
\end{equation}
where $\Tilde \omega=\omega_s-\omega$, $\Delta=\omega_s-\omega_0$, and
\begin{eqnarray}
  |G_1(\Tilde \omega)|^2 &=&\frac{\Upsilon(\Tilde \omega)+(\kappa\gamma_0-\frac{\gamma_0\gamma_1}{4})^2}{\Upsilon(\Tilde \omega)+(\kappa\gamma_0+\frac{\gamma_0\gamma_1}{4})^2}\nonumber\\
 |G_2(\Tilde \omega)|^2 &=&  \frac{\kappa\gamma_1\gamma_0^2}{\Upsilon(\Tilde \omega)+(\kappa\gamma_0+\frac{\gamma_0\gamma_1}{4})^2}
\end{eqnarray}
with $\Upsilon(\Tilde \omega)=(\frac{{\gamma_1}^2}{4}+\Tilde \omega^2)(\Tilde \omega-\Delta)^2+\frac{\gamma_0^2\Tilde \omega^2}{4}-2\kappa\gamma_0\Tilde \omega(\Tilde \omega-\Delta)$.

Although the total output spectrum is flat as given in (\ref{72-5}) due to the passivity properties of the system~\cite{guofengzhang2013}, we can apply a coherent probing field whose strength is much higher than the quantum white noise and thus the spectrum $|G_1(\Tilde \omega)|^2$ can be observed so as to calculate the spectrum $|G_2(\Tilde \omega)|^2$ which reflects the influence of the ancillary system on the output field.

\begin{figure}
  \includegraphics[width=8.5cm]{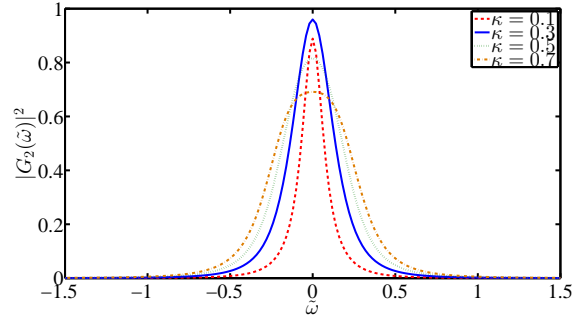}\\
  \caption{The spectrum $|G_2(\Tilde \omega)|^2$ varies with the coupling strength $\kappa$, where $\Delta=0, \gamma_0=0.1$, and $\gamma_1=0.8$.}\label{kappa}
\end{figure}

Fig.~\ref{kappa} shows the power spectral density $|G_2(\Tilde \omega)|^2$ varying with the coupling strength $\kappa$. Here, we assume there is no detuning $\Delta=0$ and the principal system damping rate $\gamma_1=0.8$ is much higher than that with respect to the white noise $\gamma_0=0.1$. As the coupling strength $\kappa$ is increased, the amplitude of the noise at the system working frequency is decreased while the bandwidth of the spectrum is broader, which means the coupling strength $\kappa$ can affect both the noise amplitude and the bandwidth.

\begin{figure}
  \includegraphics[width=8.5cm]{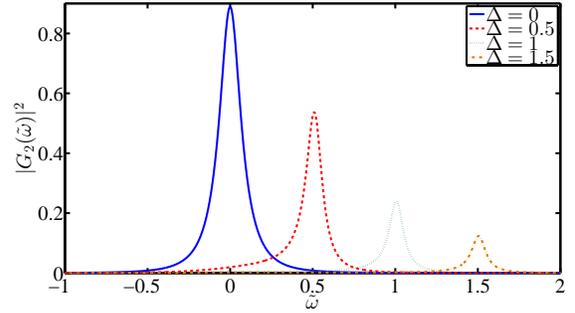}\\
  \caption{The shape of $|G_2(\Tilde \omega)|^2$ varies with $\Delta$, where $\kappa=0.1, \gamma_0=0.1$, and $\gamma_1=0.8$.}\label{delta}
\end{figure}

The power spectral density $|G_2(\Tilde \omega)|^2$ varying with the detuning $\Delta$ is plotted in Fig.~\ref{delta} with parameters $\kappa=0.1, \gamma_0=0.1$, and $\gamma_1=0.8$. When there is no detuning as given by the blue line, the noise is strong at the system frequency. As the detuning is increased via decreasing the angular frequency of the ancillary system, the spectrum $|G_2(\Tilde \omega)|^2$ is driven away from the system frequency, whose amplitude is decreased as well. This illustrates that the non-Markovian effect generated by the ancillary system becomes weaker as the detuning is increased.
\begin{figure}
  \includegraphics[width=8.5cm]{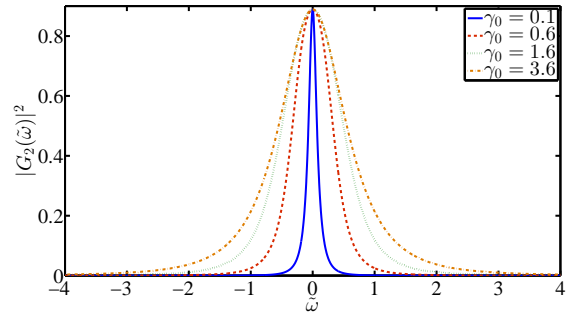}\\
  \caption{The spectrum $|G_2(\Tilde \omega)|^2$ for varying $\gamma_0$ with $\Delta=0, \kappa=0.1$, and $\gamma_1=0.8$.}\label{gamma}
\end{figure}

Fig.~\ref{gamma} shows the power spectral density $|G_2(\Tilde \omega)|^2$ can also be varied with the damping rate $\gamma_0$ with respect to the white noise. As predicted, the bandwidth of the Lorentzian spectrum is broader as $\gamma_0$ increases, varying spectrum $|G_2(\Tilde \omega)|^2$ with respect to  $\gamma_0$ gives the same result, which indicates that $\gamma_0$ determines the bandwidth of the noise spectrum.

\section{Conclusion}\label{sec6}

In this paper a Markovian representation approach to modelling a non-Markovian quantum systems compatible with quantum filtering theory has been presented.
The ancillary system of this model plays the role of the internal mode of the environment, which can convert white noise to Lorentzian noise, resulting in non-Markovian dynamics of the principal system. In addition, the quantum filter has been derived for the non-Markovian quantum system. Simulation results show the quantum filter can estimate the state of the non-Markovian quantum system.
Since the total system is expressed in an extended Hilbert space, this is equivalent to using a Markovian network to describe both the non-Markovian system and its environment. For future work, such an approach could be extended to represent a non-Markovian quantum system with arbitrary colored noise by designing the structure of the Markovian network such that a more general quantum filter can be constructed for the non-Markovian quantum systems.


%\section*{Acknowledgment}



%\bibliographystyle{IEEEtran}

\end{document}